\documentclass[11pt]{article}
\usepackage{fa2023}
\usepackage{amsmath}
\usepackage{cite}
\usepackage{url}
\usepackage{graphicx}
\usepackage{color}
\usepackage{siunitx}
\usepackage[utf8]{inputenc}
\usepackage{multirow}
\usepackage{multicol}
\usepackage{booktabs}
\usepackage{threeparttable}
\usepackage{balance}
\usepackage{adjustbox}
\usepackage[caption=false,font=footnotesize]{subfig}

\title{Data-driven 3D Room Geometry Inference with a Linear Loudspeaker Array and a Single Microphone}

\multauthor
{Cagdas Tuna$^{1*}$ \hspace{1cm} Altan Akat$^1$ \hspace{1cm} H. Nazim Bicer$^1$} { \bfseries{Andreas Walther$^1$ \hspace{1cm} Emanu\"el A. P. Habets$^2$}\thanks{$^\dag$A joint institution of the Friedrich-Alexander-Universit\"{a}t Erlangen-N\"{u}rnberg (FAU) and Fraunhofer IIS.}\\
  $^1$ Fraunhofer Institute for Integrated Circuits IIS, Erlangen, Germany\\
$^2$  International Audio Laboratories Erlangen$^\dag$, Germany
\correspondingauthor{cagdas.tuna@iis.fraunhofer.de}{Cagdas Tuna et al.}
}

\begin{document}

\maketitle
\begin{abstract}
Knowing the room geometry may be very beneficial for many audio applications, including sound reproduction, acoustic scene analysis, and sound source localization. Room geometry inference (RGI) deals with the problem of reflector localization (RL) based on a set of room impulse responses (RIRs). Motivated by the increasing popularity of commercially available soundbars, this article presents a data-driven 3D RGI method using RIRs measured from a linear loudspeaker array to a single microphone. A convolutional recurrent neural network (CRNN) is trained using simulated RIRs in a supervised fashion for RL. The Radon transform, which is equivalent to delay-and-sum beamforming, is applied to multi-channel RIRs, and the resulting time-domain acoustic beamforming map is fed into the CRNN. The room geometry is inferred from the microphone position and the reflector locations estimated by the network. The results obtained using measured RIRs show that the proposed data-driven approach generalizes well to unseen RIRs and achieves an accuracy level comparable to a baseline model-driven RGI method that involves intermediate semi-supervised steps, thereby offering a unified and fully automated RGI framework.
\end{abstract}
\section{Introduction}
\label{sec:intro}
Knowledge of geometric properties of a room can be used to improve the performance of many audio applications, including speech enhancement and dereverberation \cite{naylor2010speech}, sound source localization \cite{ribeiro2010turning}, acoustic diagnosis\cite{Dilungana2022}, sound reproduction\cite{Canclini2012}, and augmented and virtual reality \cite{Remaggi2019}. Room geometry inference (RGI) deals with the problem of reflector localization (RL), which involves estimating the time of arrivals (TOAs) of the direct path and the wall reflections from the room impulse responses (RIRs) recorded between loudspeakers and microphones. A class of RL techniques associates TOAs with reflectors by computing the common tangent to the ellipses whose foci correspond to the positions of the pair of microphones and loudspeakers \cite{antonacci2010geometric,   filos2010two, filos2012localization,  Remaggi2017}. 3D RGI is achieved in \cite{dokmanic2013acoustic} by using a distributed microphone array and computing the corresponding Euclidean distance matrix that yields the TOAs of first-order reflections. In \cite{Lovedee2019}, a 3D RGI method is proposed to localize reflectors in both convex and non-convex-shaped rooms via the spherical harmonic decomposition of spatial RIRs.  Another class of RL methods involves the disambiguation of TOAs over time-domain polar TOA/DOA (direction-of-arrival) maps produced using multi-channel RIRs \cite{Torres2013, Kim2017, Remaggi2018, ElBaba2017, ElBaba2018}. The distinct peaks emerging on these maps are associated with the temporal evolution of the direct path and the following acoustic reflections from boundary surfaces.

Many RL methods may only operate with a setup that has the same dimensions as the scenario under consideration (e.g., circular/planar arrays for 2D RL and spherical arrays for 3D RL), which would otherwise yield realistic but inaccurate RL estimates due to the geometrical ambiguity caused by the lower-dimensional array. Few RL methods exist in the literature to tackle the limited spatial diversity arising from the array. Using a compact circular microphone array, 3D RGI is achieved in \cite{Ribeiro2012} by least-squares fitting with $l_1$-regularization applied on real-world RIRs based on a broad set of synthetically generated 3D reflections. 3D RL is realized by using bi-circular microphone arrays in \cite{Remaggi2017, Remaggi2018} with the prior assumption that the array is positioned closer to the floor to tackle the up/down ambiguity. Based on the acoustic reciprocity principle \cite{Kinsler1999}, a 3D RGI method using a 2D rectangular loudspeaker array placed around a video screen and a single microphone is proposed in \cite{ElBaba2018}, where the front-back ambiguity is circumvented by assuming the array is positioned near one of the walls, and TOAs are associated with the reflectors using linear Radon transform maps \cite{ElBaba2017}.   
 \begin{figure*}[!t]
	\centerline{\framebox{
	\includegraphics[width=1.55\columnwidth]{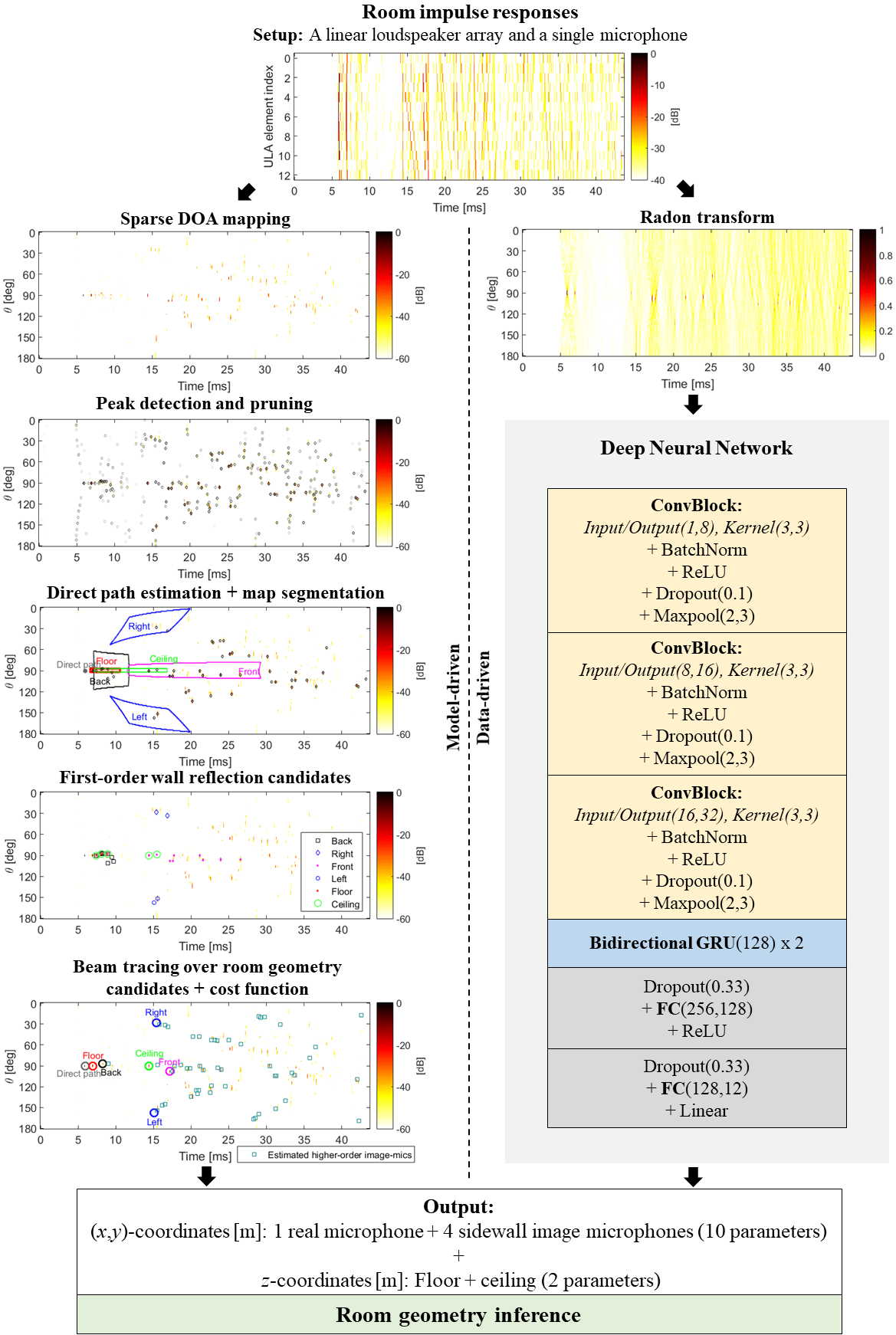}}}
	\caption{Comparison between the baseline model-driven method \cite{Tuna2020} and the proposed data-driven method using a CRNN architecture.} 
	\label{fig:block_diag}
\end{figure*}

Motivated by the growing popularity of soundbars among home entertainment systems, we have proposed a 3D RGI method using a linear loudspeaker array and a single omnidirectional microphone in \cite{Tuna2020} as the first attempt in the literature to tackle this difficult scenario with spatial diversity reduced in two dimensions due to using a 1D linear array. As illustrated in Fig.\,\ref{fig:block_diag}, this model-driven method involves multiple steps, including the generation of a computationally demanding sparse DOA map followed by a peak detection and pruning procedure, the segmentation of the sparse DOA map into six bounded regions for the identification of potential peaks associated with the first-order wall reflections, and finally, RGI using a cost function that measures the agreement between the higher-order reflections estimated via beam tracing \cite{Foco2003, Antonacci2008, Markovic2010} given a room geometry candidate and the peaks on the sparse DOA map. The model-driven approach requires coarse prior knowledge of room boundaries (i.e., pre-defined constraints on wall dimensions and orientations), for instance, to be given by the consumer in a commercial setting, along with the tuning of several parameters in other steps.

In this paper, we propose a data-driven method for 3D RGI with a linear loudspeaker array and a single omnidirectional microphone, where the microphone position and the first-order wall reflections are directly estimated from the computationally inexpensive Radon transform (RT) map \cite{Rettberg2015, ElBaba2017} being fed into a convolutional recurrent neural network (CRNN) as the input feature. As shown in Fig.\,\ref{fig:block_diag}, the proposed data-driven method eliminates the intermediate steps required in the prior model-driven approach, providing a fully automated RGI system. The deep neural network (DNN) is trained with RT maps generated from RIRs simulated in empty rooms with walls with varying absorption coefficients. The proposed data-driven 3D RGI approach is tested with both simulated and measured data and its performance is compared to the baseline model-driven method \cite{Tuna2020}.

\section{Separable Room Geometry}
\label{sec:setup}
The same setup as in the model-driven RGI method \cite{Tuna2020} is also considered here. To combat the geometrical ambiguity, a room is assumed to have a separable geometry with flat side walls of equal height that are perpendicular to the floor and ceiling. The linear loudspeaker array is positioned parallel to the floor and at the same height as the single omnidirectional microphone, further decreasing the geometrical ambiguity. 
Based on the acoustic reciprocity principle \cite{Kinsler1999}, the image-microphone positions associated with side walls lie on the same plane as the real microphone. This reduces down to a ``front-back" ambiguity, where it is still uncertain whether the image microphones are in the front or back of the array. The separable room geometry allows the individual treatment of the first-order floor and ceiling reflections, as they are located on a parallel line passing through the real microphone position perpendicular to the plane in which real and side-wall image microphones lie. This leads to an ``up-down" ambiguity between floor and ceiling. 

Without loss of generality, a uniform linear array (ULA) of $M$~loudspeakers is aligned with the $x$-axis of a reference 3D coordinate system (i.e., $m$-th loudspeaker is at $\mathbf{s}_m = [x_m, 0, 0]^T$) with the array center also coinciding with the origin. Let $\mathbf{r}$ be a point in 3D space whose projection on the 2D polar-coordinate space is given by the pair $(\rho,\theta)$, where the radial distance $\rho = \Vert \mathbf{r} \Vert$ and the polar angle $\theta = \arccos( \langle\mathbf{r},\mathbf{u}\rangle /\Vert\mathbf{r}\Vert) \in [0^{\circ}, 180^{\circ}]$ with the unit vector $\mathbf{u}=[1,0,0]^T$. Assuming that the microphone is positioned at $\mathbf{r}_o$ in front of the ULA, its Cartesian coordinates are then given by its polar-coordinate pair $({\rho}_o,{\theta}_o)$ as $\mathbf{r}_o = [\rho_o \cos\theta_o, \rho_o \sin\theta_o, 0]^T$. Using geometrical acoustics and the reciprocity principle, a specular reflection from a wall may be considered as an acoustic path stemming from the first-order image microphone positioned at $\mathbf{r}_o'$. Given a wall, its normal vector $\mathbf{v}$ and the distance to the origin $d$ can be computed using $\mathbf{r}_o$ and $\mathbf{r}_o'$:
\begin{equation}
\label{eq:normal_from_image_pos}
\mathbf{v} = \frac{	\mathbf{r}_o-\mathbf{r}_o' }{\Vert\mathbf{r}_o-\mathbf{r}_o'\Vert} \qquad \text{and} \qquad d = -\frac{1}{2}\mathbf{v}^T(\mathbf{r}_o+\mathbf{r}_o').
\end{equation}
A pair $(\rho,\theta)$ describing a first-order side-wall reflection may correspond to two points in 3D due to front-back ambiguity: $\mathbf{r}_{o',y^{\pm}} = [\rho \cos\theta, \pm\rho \sin\theta, 0]^T$, resulting in a mapping from the polar coordinates to the wall parameters as
\begin{equation}
    (\rho,\theta)\stackrel{\mathbf{r}'_{o,y^\pm}}{\longleftrightarrow} (\mathbf{v}_{y^\pm},{d}_{y^\pm}).
\end{equation} 
Assuming a separable room geometry, the image microphones associated with the first-order floor and ceiling reflections have the same \mbox{$x$--$y$} coordinates as $\mathbf{r}_o$ with up-down ambiguity: $\mathbf{r}_{o',z^{\pm}} = [\rho_o \cos\theta_o, \rho_o \sin\theta_o, \pm\sqrt{\rho^2-\rho_o^2}]^T$. Using the known unit vectors, $ \mathbf{v}_{\mathrm{floor}} =[0,0,1]^T$ and $\mathbf{v}_{\mathrm{ceiling}}=[0,0,-1]^T$ for floor and ceiling, respectively, the distance to the origin computed via (\ref{eq:normal_from_image_pos}) yields $d_z = \sqrt{\rho^2-\rho_o^2}/2$ for $\mathbf{r}_{o',z^{\pm}}$.

\section{Proposed Method}
\label{sec:method}

This section presents the deep learning framework for the proposed data-driven 3D RGI method. Let $h_m(n)$ denote an RIR recorded from the $m$-th loudspeaker in the array to the single microphone, where a synchronized setup is considered, i.e., the initial delay in $h_m(n)$ is associated with the direct path propagation. The Radon transform, equivalent to a full-band delay-and-sum beamformer \cite{Rettberg2015, Ryan1997, ElBaba2017}, is given by
\begin{equation}
    R(n, \theta) = \sum_{m = 0}^{M-1}   d^{(m)}_{n, \theta} \, h_m^+\left(n - \left \lfloor \Delta^{(m)}_{n, \theta}  \right\rceil \right),
\end{equation}
where $ d^{(m)}_{n, \theta}  = \Vert\mathbf{r}_{\rho_n, \theta} - \mathbf{s}_m\Vert$ is the propagation path distance to the point $\mathbf{r}_{n, \theta} = [\rho_n \cos\theta, \rho_n \sin\theta, 0]^T $ from the $m$-th loudspeaker $\mathbf{s}_m$, $\Delta^{(m)}_{n, \theta} = \frac{f_s}{c} \left (\rho_n - d^{(m)}_{n, \theta} \right )$ is the associated time delay in samples, and $f_s$ and $c$ denote the sampling frequency and the speed of sound, respectively. Only the positive side of the RIR denoted by $h_m^+(n)$ is considered to exclude any redundant peaks due to the loudspeaker impulse response, which may otherwise introduce spurious peaks on the generated RT map. Linear interpolation is applied in the actual RT implementation to prevent potential artifacts due to the rounding operation $\lfloor.\rceil$. The resulting map is fed to the DNN after normalization by the maximum value.

The CRNN architecture used in the proposed data-driven RGI method is also illustrated in Fig.\,\ref{fig:block_diag}. The microphone position and the four image-microphone positions corresponding to the first-order side-wall reflections in 2D Cartesian $xy$-coordinates, as well as the $z$-coordinates of the image-microphone positions for floor and ceiling, are estimated from the RT map by the DNN. The room geometry is then inferred from the microphone and first-order image-microphone position estimates using (\ref{eq:normal_from_image_pos}). 

The output of the convolutive layers is reshaped across the angular direction while keeping the time resolution unchanged and then fed into the two-layer bidirectional gated recurrent unit (GRU). The GRU output is fed into the successive fully connected (FC) layers. The rectified linear unit (ReLU) function is used as the activation function at the output of convolutive layers. The batch normalization is performed before ReLU, while dropout and max pooling are applied after ReLU. The dropout is also applied at the input of both FC layers, where the first layer uses ReLU as the activation function, and the second layer has linear activation to accommodate for negative-valued $xy$ coordinates.

The loss function used to train the network is based on the Euclidean distance between the actual and estimated real- and image-microphone positions and is defined as
\begin{equation}
    \textrm{Loss} = \frac{1}{7} \sum_{w=0}^{6} \epsilon_w,
\end{equation}
where for the real microphone ($w = 0)$ and four side walls ($w \in \{1,2,3,4\}$),
\begin{equation}
    \epsilon_w = \sqrt{(x_w - \hat{x}_w)^2 + (y_w - \hat{y}_w)^2},
\end{equation}
and for floor and ceiling ($w \in \{5,6\}$), 
\begin{equation}
    \epsilon_w = \left|z_w - \hat{z}_w \right|
\end{equation}
with ($x_w, y_w, z_w$), and ($\hat{x}_w, \hat{y}_w, \hat{z}_w$) denoting the ground-truth and estimated coordinates, respectively. The batch size is chosen to be $50$, and the network optimizer is AdamW\cite{Loshchilov2019} with a learning rate of $5\times10^{-4}$ and a weight decay rate of $10^{-2}$. Early stopping is used to prevent overfitting. 
\begin{table}[!b]	
\centering
	\caption{Wall constraints in \cite{Tuna2020} for small rooms. The minimum distance $d_w^{\min}$ and maximum distance $d_w^{\max}$ are measured from the origin to wall $w$.}
	\label{tab:rgi_constraint}
	\begin{tabular}{lcc}
 \\
		\toprule
		\multicolumn{1}{l}{Wall}& $d_{w}^{\min}$ & $d_{w}^{\max}$ \\  \midrule
		Back & $0.2$ m & $1.0$ m   \\ 
		Right  & $1.5$ m & $3.0$ m    \\ 
		Front & $3.0$ m & $6.0$ m   \\ 
		Left  &  $1.5$ m & $3.0$ m  \\ 
		Floor  & $0.5$ m & $1.5$ m \\ 
		Ceiling  &  $0.7$ m & $2.7$ m  \\ 
		\bottomrule
	\end{tabular} 
\end{table}

\section{Performance Evaluation}
\label{sec:evaluation}

RIRs were simulated using the \textit{Pyroomacoustics}\cite{Dokmanic2018} software package with the image method \cite{allen1979image} up to and including fifth-order reflections between a ULA of 13 loudspeakers spaced by $6$ cm and a single microphone, which were all assumed to be omnidirectional. The wall absorption coefficients were randomly selected between $0.1$ and $0.9$. The same wall constraints as in \cite{Tuna2020} chosen for the generation of bounded regions on the DOA map for small rooms were also used here in RIR simulations as listed in Table \ref{tab:rgi_constraint}, where the walls were named with respect to the ULA placement in the room and the ULA was positioned closer to the back wall, considering a typical living room setting with a soundbar. Also in line with \cite{Tuna2020}, the side walls were allowed to have an orientation angle with up to a deviation of $15^\circ$ from a shoe-box room model, and the minimum room height was set to $H_{\min} = 2.2$ m with the presumption that ULA was closer to the floor to tackle up-down ambiguity (which implies the minimum distance for the ceiling is $d_w^{\min} > 1.1$ m in simulated rooms). The microphone was randomly positioned at a minimum distance of $0.5$ m from the side walls and the ULA.  $50,000$ rooms were randomly generated based on the room constraints for network training. An additional $5,000$ rooms were generated each for validation and for testing. For RIR simulations, the sampling frequency was $f_s = 48$ kHz, and the speed of sound was assumed to be $c$ = 343 m/s. Simulated RIRs were then low-pass filtered at $f_c = 20$ kHz. The RT maps were generated up to the propagation distance of $\rho_{\max} = 15$ m (or equivalently $t_{\max} \approx 43.73$ ms,  $N = \lfloor f_s(\rho_{\max} /c)\rceil = 2099$ samples) with the angular grid of $[0^{\circ},180^{\circ}]$ with $1^\circ$ resolution.

The performance metrics used for the evaluation of wall estimation accuracy are the distance and orientation errors defined as\cite{Tuna2020} 
\begin{equation}
\epsilon_{{w},d} = \left| d_{w} - \hat{d}_{w}\right|\quad \text{and}\quad \epsilon_{{w},\theta} = \arccos \langle \mathbf{v}_{w}, \hat{\mathbf{v}}_{w}\rangle,
\label{eq: eps}
\end{equation}
which measure the deviation of the estimated wall parameters $(\hat{\mathbf{v}}_{w}, \hat{d}_{w})$ from the ground truth values $(\mathbf{v}_{w}, d_{w})$. As in \cite{Tuna2020}, RGI accuracy is evaluated based on the root MSE computed over the individual wall estimate errors:
\begin{equation}
E_d = \sqrt{\frac{1}{6}\sum_{w=1}^6 \epsilon_{w,d}^2}, \quad
E_\theta = \sqrt{\frac{1}{4}\sum_{w=1}^4 \epsilon_{w,\theta}^2}.
\end{equation}
Please note that under the separable room geometry assumption, $\epsilon_{{w},\theta} = 0$ for floor and ceiling, such that $E_\theta$ can be computed considering only the side walls.

\begin{table}[!b]	
\centering
\begin{threeparttable}
	\caption{Simulated test data: The DNN performance evaluation.}
	\label{tab:rgi_sim}
	\begin{tabular}{lcc}
		\toprule
		\multicolumn{1}{l}{Wall}& Distance err. [cm] & Orientation err. [$^\circ$]\\  \midrule
		Back & $ 6.503 \pm 5.021$ & $ 1.508 \pm  1.308$    \\ 
		Right  & \hspace{-2mm}$10.516 \pm 9.277$ & $ 2.042 \pm 1.903 $     \\ 
		Front & $11.354 \pm 11.216$ & $ 2.809 \pm  2.867$    \\ 
		Left  & \hspace{-2mm}$10.472 \pm  9.394$ & $ 1.944 \pm 1.753$   \\ 
		Floor  & $  3.505 \pm 3.142$ & -   \\ 
		Ceiling  & $ 6.180 \pm 7.057$ & -   \\ \hline
		Room & \hspace{-2mm}$10.471\pm 5.309$ & $ 2.582 \pm  1.421$ \\
		\bottomrule
	\end{tabular} 
		    \begin{tablenotes}
      \footnotesize
      \item Mean$\pm$std (standard deviation)  computed over 5000 rooms.
    \end{tablenotes}
\end{threeparttable}
\end{table}

 \begin{table*}[!t]	
\centering	
 \begin{threeparttable}
	\caption{Measured data: Performance comparison. Mean$\pm$std (standard deviation) are computed over 27 measurement positions from the three rooms shown in Fig.~\ref{fig:rgi_exp}.}
	\label{tab:rgi_exp}
	\begin{tabular}{lcccc}
		\toprule
		& \multicolumn{2}{c}{Distance error [cm]} &\multicolumn{2}{c}{Orientation error [$^\circ$]} \\
		\multicolumn{1}{l}{Wall}& Model-driven \cite{Tuna2020} & Data-driven & Model-driven \cite{Tuna2020} & Data-driven\\  \midrule 
		Back & $ 14.396 \pm 13.114$ & \hspace{2mm}$7.704 \pm 11.702$ & $0.983 \pm 1.048$ & $1.338 \pm1.095$   \\ 
		Right  & $4.117\pm 6.090$ & \hspace{-2mm}$14.897 \pm 4.578 $  & $1.510 \pm 1.867$ & $1.502 \pm 1.037$   \\ 
		Front & $15.557 \pm 32.352$ & $11.736 \pm 19.875$ & $0.780 \pm 0.585$ & $1.647 \pm1.319$   \\ 
		Left  & $  5.526 \pm 4.387$ & $ 5.627 \pm 4.627$ & $1.835\pm1.628$ & $1.243 \pm 1.574$  \\ 
		Floor  & $  1.000\pm 2.095$ & $ 4.765 \pm 2.196$ & - & -  \\ 
		Ceiling  & $ 2.430\pm 4.464$ & $ 7.887 \pm 6.855$ & - & -   \\ \midrule
		Room & $  12.007\pm 12.410$ & $ 11.913 \pm 7.191$ & $1.657 \pm 0.958$ & $1.758 \pm 0.779$ \\
		\bottomrule 
	\end{tabular}
  \end{threeparttable}
\end{table*}

\begin{figure}[!t]
    \begin{adjustbox}{varwidth=\columnwidth,fbox,center}
	\centering
\subfloat[Small office: Configuration A ($\mathrm{RT}_{60} = 0.57\,\mathrm{s}$)]{%
     \centering
      \includegraphics[width=.95\columnwidth]{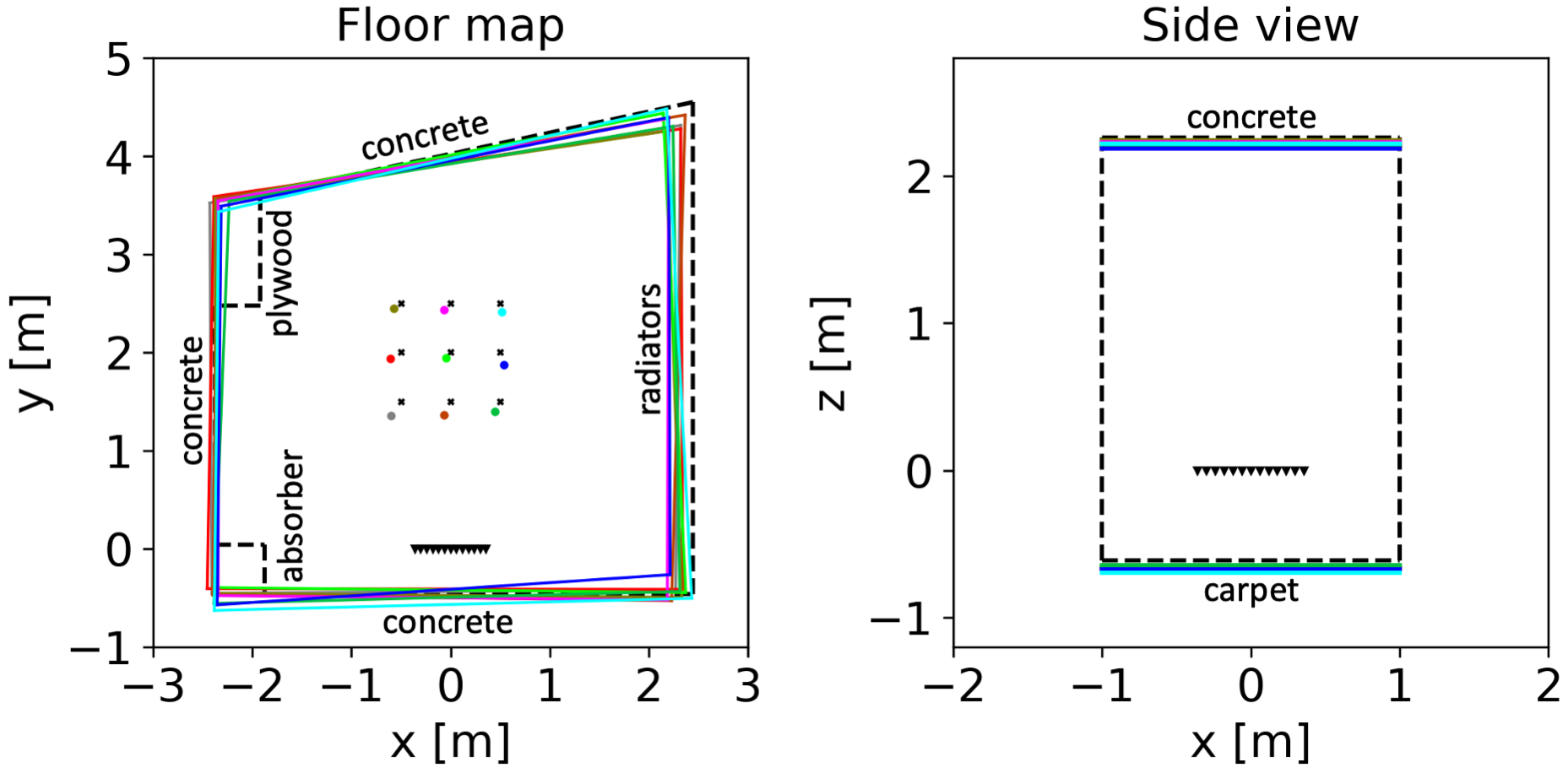}	
     }

     \subfloat[Small office: Configuration B ($\mathrm{RT}_{60} = 0.57\,\mathrm{s}$)]{%
     \centering
       \includegraphics[width=.95\columnwidth]{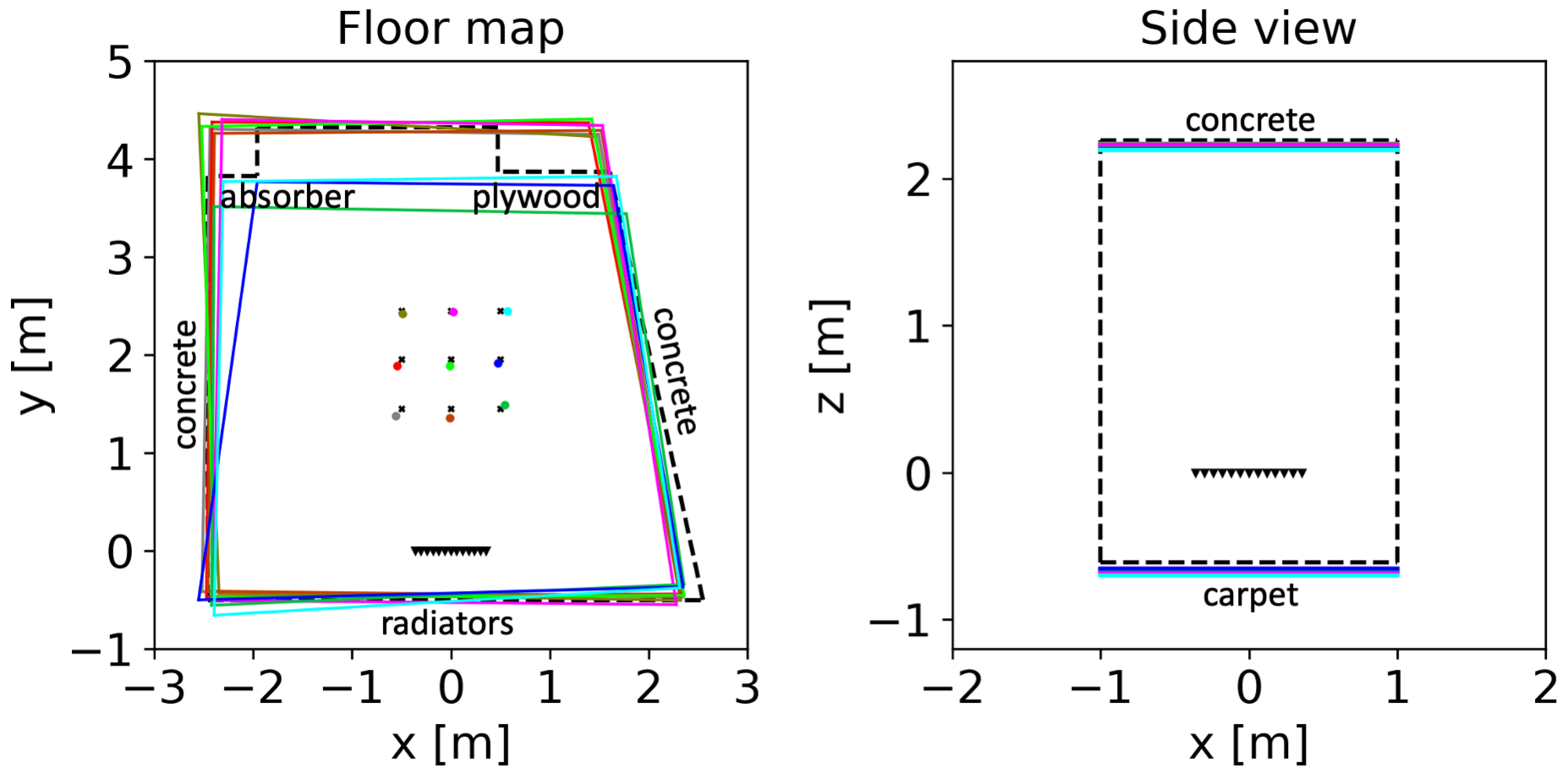}	
     }
     
        \subfloat[Laboratory room  ($\mathrm{RT}_{60} = 0.7\,\mathrm{s}$)]{%
     \centering
       \includegraphics[width=.95\columnwidth]{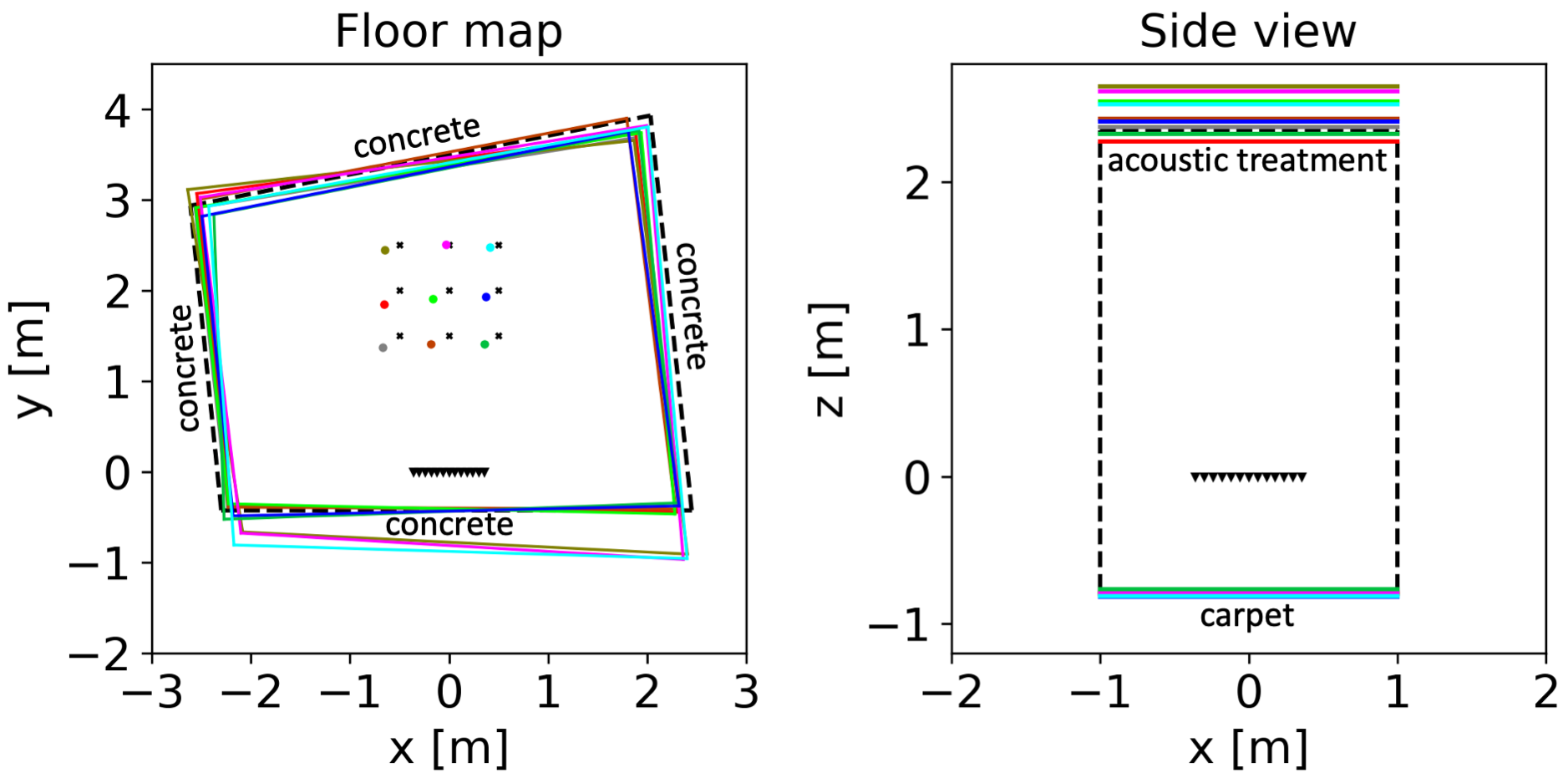}	
     }
     \end{adjustbox}
          \caption{RGI results with measured RIRs: Inferred floor maps, floor/ceiling heights, and the estimated microphone positions. Dashed lines indicate the ground truth.}
     \label{fig:rgi_exp}
 \end{figure}

The DNN performance is first tested with the simulated test data as presented in Table\,\ref{tab:rgi_sim}. The RGI mean-error values $\mu(E_d) < 11$ cm and $\mu(E_\theta) < 3^\circ$ indicate that the proposed data-driven approach shows promising performance despite the geometrical ambiguities. The highest accuracy for distance estimation was achieved for the floor, followed by the ceiling and the back wall. This can be explained by the room simulation setup with the ULA, which was positioned closer to the back wall than the other side walls and parallel to the floor and ceiling (i.e., not tilted). The highest errors for wall distance and orientation were reached for the front wall, which may be expected as it was farther from the ULA.  

The proposed data-driven approach was also tested with the real RIRs measured in a small office under two configurations and a laboratory room. Please refer to \cite{Tuna2020} for further details on the measurement campaign. The performance comparison between the model- and data-driven RGI methods is presented in Table\,\ref{tab:rgi_exp}, and the inferred floor maps and room heights are plotted in Fig.\,\ref{fig:rgi_exp} for all three setups. Despite training with only simulated RIRs, the data-driven method yields RGI mean-error values $\mu(E_d)$ and $\mu(E_\theta)$ comparable to those achieved by the model-driven technique when tested with unseen measured RIRs. The source of wall estimation errors was similar to the ones observed in the baseline method. The impact of violating flat-wall assumption can be seen in the small office. For instance, some wall estimates were aligned with the plywood cabinet and/or absorber, while the radiators in Configuration~A were detected as the right wall. The ceiling distance estimation was perturbed in the laboratory room by the acoustic treatment and the air ventilation duct going through just below the ceiling. The distance errors for the back wall were visibly larger at three microphone positions due to the associated peaks on the RT maps having very weak amplitudes caused by the loudspeaker directivity. In addition, the RGI performance is very close to the one achieved with the simulated RIRs, meaning that the proposed data-driven approach generalizes well to unseen measured RIRs. Preliminary trials with the DNN architecture design have indicated that this was achieved particularly by using dropout in convolutive layers, creating a regularization effect for the measured data at the cost of a slightly increased error for the simulated data.

The proposed approach offers a more unified solution using an input feature map that is relatively cheap to compute. As once the DNN is trained, it directly outputs the estimates of the real- and image-microphone positions, bypassing all the steps requiring a variety of parameter tuning in the model-driven approach,  including sparse DOA map generation, peak detection and pruning, and map segmentation demanding a user input for wall constraints. 
Another main issue with the model-driven method is the cost function used for RGI, which depends on the amplitude and the position of higher-order reflections on the sparse DOA map, making it vulnerable in non-empty rooms with furniture or in the case of diffuse reflections. Although the current DNN was only trained with data from simulated empty rooms with perfectly flat walls, a future data-centric approach to improve the robustness of the data-driven RGI may involve adding real data measured in such room conditions.

\section{Conclusion}
A data-driven method for 3D RGI with a linear loudspeaker array and a single microphone in convex-shaped rooms has been proposed. It has been shown that although the DNN has only been trained with simulated data, it has achieved a level of accuracy comparable to the baseline model-driven method when tested with real-world data. It has also been pointed out that the proposed data-driven approach eliminates the need for all the intermediate steps involved in the model-driven method and instead directly estimates the positions of the microphone and the wall reflections in polar coordinates from the RT map. Future work includes extending the training dataset by adding real-world data measured in non-ideal scenarios, such as rooms with furniture and/or diffuse reflections, and expanding the operation into larger rooms to increase the generalizability of the data-driven RGI method.   

\section{Acknowledgments}
Parts of this work have been funded by the Free State of Bavaria in the DSAI project.

\balance
\bibliography{refs}

\end{document}